\documentclass[12pt]{article}
\usepackage{amsmath}
\usepackage{amssymb}
\usepackage{amstext}
\usepackage{graphicx}
\usepackage{color}
\usepackage{cite}

\renewcommand{\arraystretch}{1.7}

\newcommand\ZZ{Ziolkowski }
\newcommand\W{\ell_0}
\newcommand\FF{{ \cal F }}
\newcommand\HH{{ \cal H }}

\newcommand\WW{{ W }}
\newcommand\CE{\textrm{CE}}
\newcommand\CM{\textrm{CM}}
\newcommand\real{\mathbb{R}}
\newcommand\complex{\mathbb{C}}
\newcommand\J{\mathcal{J}}
\newcommand\wt[1]{\widetilde{#1}}

\newcommand\sqrr{ \sqrt{ k^2-(\frac{\omega}{c}  )^2 }  }
\newcommand\sqr{ \sqrt{ (\frac{\omega}{c}  )^2 - k^2 }  }
\newcommand\GAM{ \Gamma }
\newcommand\EE{ \widehat{ \cal E  }}
\newcommand\params{$(\Lambda, \Psi_1,\Psi_2, \Phi,\Xi  )\,$}

\newcommand\THREEVEC[3]{	\left(\!	\begin{array}{c}
											#1 \\  
											#2 \\
											#3 
										\end{array}\!\right) }

\newcommand\bfx[1]{(t,{\mathbf x}_{#1})}
\newcommand\alp[1]{{\alpha_{#1}}}
\newcommand\bta[1]{{\beta_{#1}}}
\newcommand\bfalp{{\boldsymbol\alpha}}
\newcommand\bfbta{{\boldsymbol\beta}}

\newcommand\POWERPULSEPLOT{
\begin{figure}[!h]
		\centering
		\includegraphics{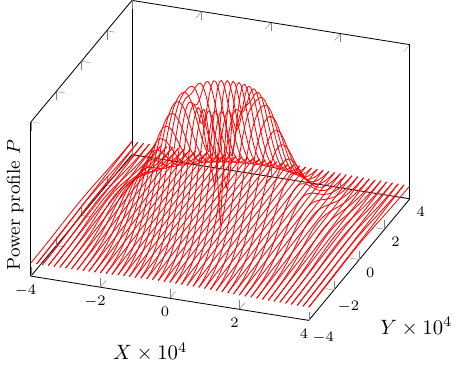}
	\caption{Power profile $P$  of the $(\text{CM},1)$ laser pulse at $Z=0,\,T=0$  with parameters $\{\Lambda=600, \Psi_{1}=1, \Psi_{2}=1000, \Phi=0.001, \Xi=1\}$}
	\label{LPPD}
\end{figure} 
}

\newcommand\LASERSPACECURVES{
\begin{figure*}[!ht]
		\centering
		\includegraphics{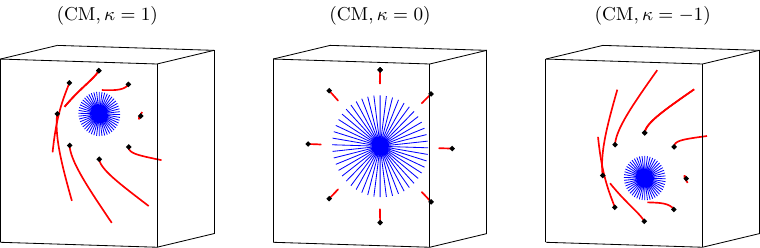}
	\caption{Three-dimensional spacecurves for particles subject to an incident $(\text{CM},1)$ laser pulse (left), $(\text{CM},0)$ laser pulse (centre) and $(\text{CM},-1)$ laser pulse (right) with parameters $\{\Lambda=600, \Psi_{1}=1, \Psi_{2}=1000, \Phi=0.001, \Xi=1\}$. Each particle has initial velocity $\{\dot{R}(0)=0, \dot{\theta}(0)=0, \dot{Z}(0)=\frac{1}{200}\}$. The shaded  circular disc region indicates the initial  spot size ($R=10000$ for $(\text{CM},\pm 1)$ laser pulses and $R=20000$ for a $(\text{CM},0)$ laser pulse) relative to the black markers on the spacecurves that denote the initial positions of the charged test particles.}
	\label{DTP}
\end{figure*} 
}

\newcommand\LASERSKE{
\begin{figure*}[htbp]
		\centering
		\includegraphics[width=0.98\textwidth]{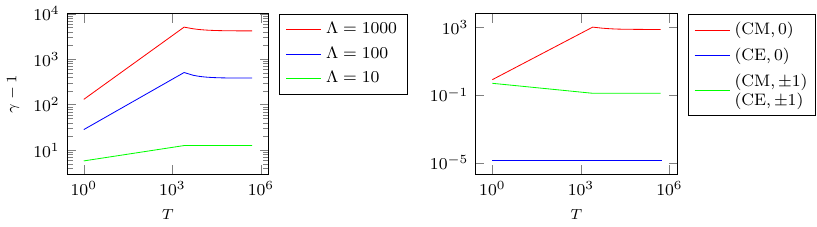}
	\caption{Specific kinetic energy transfer $\gamma-1$ to any charged test particle. On the left, the (CM,$1$) pulse has parameters $\{\Psi_{1}=1, \Psi_{2}=1000, \Phi=0.001, \Xi=1\}$. On the right, relative values of $\gamma-1$ are displayed for various $(s,\kappa)$ pulses with parameters $\{\Lambda=1, \Psi_{1}=1, \Psi_{2}=1000, \Phi=0.001, \Xi=1\}$. The differences between the energy transfers for some pulses appear indistinguishable relative to others owing to the logarithmic scales employed. In all cases, the charged particle has initial position $\{R(0)=1, \theta(0)=\frac{\pi}{2}, Z(0)=1\}$ and initial velocity $\{\dot{R}(0)=0, \dot{Z}(0)=\frac{1}{200}, \dot{\theta}(0)=0\}$.}
	\label{SKE}
\end{figure*} 
}

\newcommand\LASERTABLE{
	\begin{table*}[b]
	\centering\small
	\renewcommand{\arraystretch}{1.2}
		\begin{tabular}{|c|c|c|c|c|c|c|c|c|}
			\hline
			 $s$ & ${\;\kappa\,\;}$ & ${\mathcal{E}}$ (J) & $\Lambda$ & ${z_{rg}}$ (m)\, & ${t_{0}}$ (ps) & ${z_{w}}$ (m) &  ${r_{s}(0)}$ (m) & $I$ (W\,$\text{cm}^{-2}$)\\
			\hline
			\hline
			CE & 0 & 8.07 & 40000 & 1.44 & 3 & $ 8.99\times 10^{-4}$ & 0.018 & $ 2.646 \times 10^{12}$ \\
			\hline
			CE & $\pm 1$ & 7.69 & 1 & 2.43 & 3 & $ 8.99 \times 10^{-4}$ & 0.009 & $ 1.009 \times 10^{12}$ \\
			\hline \hline
			CM & 0 & 8.41 & $\frac{1}{20}$ & 0.899 & 2 & $ 6.00\times 10^{-4}$ & 0.012 & $ 9.304 \times 10^{11}$ \\
			\hline
			CM & $\pm 1$ & 7.69 & 1 & 1.80 & 3 & $ 8.99 \times 10^{-4}$ & 0.009 & $ 1.009 \times 10^{12}$ \\
			\hline\hline
			CM & $\pm 1$ & 76903 & 100 & 1.80 & 3 & $ 8.99\times 10^{-4}$ & 0.009 & $ 1.009\times 10^{16}$ \\
			\hline
		\end{tabular}
		\label{MKSLPC}
		\caption{Table showing SI laser characteristics for various $(s,\kappa)$ laser pulse configurations with parameters $\{\Psi_{1}=1, \Psi_{2}=1000, \Phi=0.001, \Xi=1\}$.}
	\end{table*}
}

\begin{document}
	\begin{center}
		{\Large \bf The Dynamics of Compact Laser Pulses} \\[1cm]
		{\large S. Goto$^{1,}\footnote{Current address: Department of Applied Physics and Mathematics, Graduate School of Informatics, Kyoto University, Yoshida-Honmachi, Sakyo-ku, Kyoto 606-8501, Japan }$, R. W. Tucker$^{2}$ and T. J. Walton$^{3}$} \\
	\end{center}
	{$^{1}$ {\small Institute for Molecular Science, 38 Nishigo-Naka, Myodaiji, Okazaki 444-8585, Japan}}  \\
	{$^{2}$ {\small Department of Physics, University of Lancaster and Cockcroft Institute, Daresbury Laboratory, Warrington, UK}} \\
	{$^{3}$ {\small Department of Mathematics, University of Bolton, Deane Campus, Bolton, UK}}	\\[0.6cm]
				
	{Email: {\small sgoto@ims.ac.jp, \, r.tucker@lancaster.ac.uk, \, t.walton@bolton.ac.uk}} \\

	\begin{abstract}
		We discuss the use of a class of exact finite energy solutions to the vacuum source-free Maxwell equations  as models for  multi- and single cycle laser pulses in classical interaction with relativistic charged point particles. These compact solutions are classified in terms of their \textit{chiral} content and their influence on particular charge configurations in space. The results of such classical interactions motivate a phenomenological quantum description of a propagating laser pulse in a medium in terms of an \textit{effective} quantum Hamiltonian.
	\end{abstract}

	\quad \\[-0.1cm]
	
					
	\quad \\[0.3cm]

\section{Introduction}
Advances in laser technology have made possible the exploration of physical processes on unprecedented temporal and spatial scales. They have also opened up new possibilities for accelerating charged particles using laser-matter interactions. Multi- and single cycle high intensity ($10^{10}- 10^{15}\,\text{W}\text{cm}^{-2}$) laser pulses can be produced using Q-switching or mode-locking techniques \cite{brabec2000intense}. Pulses of even higher intensity ($\sim 10^{21} \,\text{W}\text{cm}^{-2}$) could accelerate charged particles such as electrons to relativistic speeds where radiation reaction and quantum effects may influence their dynamics \cite{yoffe2015}. Lower intensity pulses have also been used as diagnostic tools for exploring the structure of plasmas in various states \cite{buck2011real, matlis2006snapshots}. In order to interpret experimental data involving classical laser interactions with both charged and neutral matter, theoretical models \cite{mackenroth2010determining, tsung2002generation, ziolkowski1991collective, ziolkowski2006metamaterial} rely crucially on parameterizations of the electromagnetic fields in laser pulses, particularly in situations where traditional formulations using monochromatic or paraxial-beam approximations have limitations \cite{liu2011exact, harvey2011symmetry, terranova2014particle}. Such theoretical models may not be compact in all spatial dimensions and the role of a laser pulse as a classical  probe is further limited by the scales that it is designed to resolve. Pulse shape design characteristics are often guided by simulations of laser-matter interactions which incorporate the known laws of physics of relevance at such scales. 
For example, the intense experimental activity currently exploring the electromagnetic properties of single-cycle laser pulses with nanoscale objects  (such as dielectric and plasmonic nanoparticles) demands efficient modelling tools that accommodate the spatial compactness of such pulses. Such tools will eventually require incorporation of quantum effects associated with these interactions in order to properly describe observations and yield practical applications involving nanostructured dielectrics and plasmonic metals \cite{raybould2015focused}. Furthermore an effective quantum description of  the laser pulse itself would offer  a new  simulation tool for designing more accurate methods of encoding quantum information. 

The quanta associated with {\it plane time-harmonic} electromagnetic fields in vacuo provide ideal discrete two-state systems (photons) that are routinely used as controllable qubits in information science.  Three-level bosonic quantum systems composed of two photons in the same spatial and temporal configuration have also been contemplated \cite{klimov2003qutrit} in attempts to construct more efficient quantum gates for quantum communication. If one regards a free propagating classical single-cycle (therefore non-time-harmonic) laser pulse in vacuo as a spatially compact classical electromagnetic configuration with definite energy ${\cal{E}}$  and temporal width $t_0$, one expects that when ${\cal{E}} t_0 \lesssim \hbar$ its dynamical evolution, in both vacuo and material media,  should be controlled by an effective quantum Hamiltonian, rather than a classical one as is done in \cite{ziolkowski1995ultrafast}. Such quantized collective states  could then be entangled with other quantized pulses or free photon states and their interaction with classical or quantized states of electrically neutral continua  (e.g. optically inhomogeneous and anisotropic dielectrics or plasmas) or charged matter (e.g. trapped ions \cite{klimov2003qutrit}) may be worthy of investigation for technological applications such as quantum computing and encryption. 

In this article we first discuss a viable methodology for parameterizing a particular class of propagating solutions to the source-free {\it classical } Maxwell equations in vacuo that offers an efficient  means to explore the  classical effects of \textit{compact} laser pulses on free electrons in dynamical regimes where quantum effects are absent. The parameterization is constructed from a remarkable class of explicit solutions of the scalar wave equation found by \ZZ \cite{ziolkowski1985exact, ziolkowski1989localized, shaarawi1989localized, donnelly1992method, donnelly1993designing} following pioneering work by Brittingham \cite{brittingham1983focus} and Synge \cite{synge1956relativity}. Solutions in this class are parameterized in terms of three real constants that are sufficient to completely determine the characteristics of any freely propagating laser pulse in full accord with Maxwell's equations in free space. They describe solutions with finite total electromagnetic energy, electromagnetic fields bounded in \textit{all} three spatial directions and experimentally distinguishable \textit{chiral configurations}. With simple analytic structures their diffractive properties can be readily calculated together with the behaviour of relativistic charged particle-pulse interactions over a broad parameter range without recourse to expensive numerical computation. Based on this behaviour and by analogy with the effective Hamiltonian theory of diatomic molecules, we are led to construct a Hilbert space on which to describe  certain \textit{quantum states} of an electromagnetic pulse undergoing unitary evolution generated by an effective phenomenological {\it quantum Hamiltonian} in a medium. Such a Hamiltonian, defined in terms of a set of parameters associated with its medium interaction, serves as an effective quantum model for the quantum pulse evolution.

\section{Parameterizing Compact Electromagnetic Laser Pulses}

If a complex scalar field $\alpha $ satisfies $\Box\,\alpha=0$  and $ \Pi_{\mu\nu}$ is any covariantly constant (degree 2) anti-symmetric tensor field on spacetime (i.e. $\Pi_{\mu\nu ;\delta }=0$ for all $\mu,\nu,\delta =0,1,2,3  $), then the complex tensor field $F_{\mu\nu}=\partial_{\mu} A_{\nu} -  \partial_{\nu} A_{\mu} $ satisfies the source-free Maxwell equations in vacuo with:
\begin{eqnarray}\label{ALP}
	A_{\nu} &=& \partial_{\gamma}\left (  \alpha \,\,\Pi_{\mu\beta  }  \right)  \, \epsilon^{\gamma\mu\beta   }{}_{\nu}\, \sqrt{ \vert  g \vert }
\end{eqnarray}
where $\vert g \vert$ is the  modulus of the determinant of the spacetime metric tensor field $g$ with components $ g_{\mu\nu } $ and $\epsilon^{\gamma\mu\beta}{}_{\nu}\,$ denotes the Levi-Civita alternating symbol. In the following, we restrict to  Minkowski spacetime, in which case the components $\Pi_{ \mu\nu }$ can be used to encode three independent Hertz vector fields and their duals\footnote{In the language of differential forms on Minkowski spacetime $A=\star d(\alpha \Pi),\,\, F=dA$ where $d\star d\,\alpha=0$, the 2-form $\Pi$ satisfies $\nabla \Pi =0$ and $\star$ denotes the Hodge map associated with $g$.}.

General solutions to  $\Box \,\alpha =0$ can be constructed by Fourier analysis. In cylindrical polar Minkowski coordinates $\{t,r,z,\theta\}$, axially symmetric solutions propagating along the $z$-axis have, for $z\geq 0$, the double integral representation $\alpha(t,r,z)=\int_{-\infty}^{\infty} \, d\omega \, e^ {-i \omega t} \wt{\alpha}(\omega,r,z)$ where:
\begin{eqnarray*}
	\wt{\alpha}(\omega,r,z) &=& \int_{0}^{ \frac{\omega}{c} }  k \, f_\omega(k) J_0(k r) e^{ \pm i z \sqr  }   \, dk  + \int_{ \frac{\omega}{c} } ^ \infty k f_\omega(k) J_0(k r) e^{ - z \sqrr  }   \, dk      
\end{eqnarray*}
in terms of the zero order Bessel function and the speed of light in vacuo $c$. 

\POWERPULSEPLOT

Conditions on the Fourier amplitudes $f_\omega(k)$ can be given so that the Hertz procedure above gives rise to real singularity free electromagnetic fields with finite total electromagnetic energy. A particularly simple class of pulses that can be generated in this way follows from the complex axi-symmetric scalar solution:
\begin{eqnarray}\label{alp}
   \alpha(t,r,z) &=& \frac{\W^2}{r^2 + ( \psi_1 + i(z-ct)  )\,(  \psi_2-i( z+ct )  ) }
\end{eqnarray}
where $\W, \psi_1,\psi_2$ are strictly positive (real) parameters with physical dimensions of length. The relative sizes of $\psi_1$ and $\psi_2$  determine  both the direction of propagation  along the $z-$axis of the dominant maximum of the pulse profile. When $\psi_{1}\gg\psi_{2}$, the dominant maximum propagates along the $z-$axis to the right. The parameter $\ell_0$ determines the magnitude of such a maximum. The structure of such solutions has been extensively studied in \cite{fedotov2007exact,feng1999spatiotemporal} in conjunction with particular choices of $\Pi_{\mu\nu}$ together with generalizations discussed in \cite{borzdov2002designing,hernandez2007localized}.

\LASERSPACECURVES

In general the six anti-symmetric tensors with components $\delta^\mu_{ [\gamma } \delta^\nu_{\sigma ]}$ in a Minkowski Cartesian coordinate system are covariantly constant and can be used to construct a complex eigen-basis of antisymmetric {\it chiral} tensors $ \Pi^{s,\,\kappa }$, with  $s \in \{\text{CE},\text{CM}\} $ and $\kappa \in \{ -1,0,1 \}$,  satisfying
\begin{eqnarray} \label{chiral}
	{\cal O}_z  \, \Pi^{s,\,\kappa } &=& \kappa\, \Pi^{s,\,\kappa }
\end{eqnarray}
where the operator ${\cal O}_z$ represents  $\theta$ rotations about the $z-$axis generated by $-i \partial_\theta $ on tensors\footnote{In terms of the Lie derivative, ${\cal O}_z=-i {\cal L }_{\partial_{\theta} }$  and
$\Pi^{ \CE,\pm 1 }=d(x\pm i y)\wedge dt,\,\, \Pi^{ \CE,0 }=dz\wedge dt, \Pi^{ \CM,\kappa } =  \star \Pi^{ \CE,\kappa }$ where $x=r \cos(\theta), y=r \sin(\theta)$}. These in turn can be used to construct a {\it complex basis} of chiral eigen-Maxwell tensor fields $F^{s,\,\kappa}_{\mu\nu}$. The index $s$ indicates that the CE (CM) chiral family contain electric (magnetic) fields that are orthogonal to the $z-$axis when $\kappa=0$. The chiral eigen-fields $F^{s,\,0}_{\mu\nu }$ inherit the axial symmetry of $\alpha(t,r,z)$ while those with $\kappa=\pm 1$ do not. The directions of electric and magnetic fields in any of these  Maxwell solutions depend on their location in the pulse and the concept of a pulse polarisation is not strictly applicable. The chiral content as defined here can be used in its place. Non-chiral (complex) pulse configurations can be constructed by superposition: $F_{\mu\nu}= \sum_s\, \sum_\kappa F^{s,\,\kappa}_{\mu\nu}  \, {\cal C}^{s,\,\kappa}$ with arbitrary constant complex coefficients ${\cal C}^{s,\,\kappa}$.

The energy, linear and angular momentum of the pulse in vacuo can be calculated from the components $T_{\mu\nu}$ of the Maxwell stress-energy  tensor $T_{\mu\nu}=-\frac{1}{4} g_{\mu\nu} \FF^{\alpha\beta}\FF_{\alpha\beta} - \FF_{\mu\alpha} \FF^{\alpha}{}_{\nu} $
where $\FF_{\mu\nu}= \text{Re}(F_{\mu\nu})$. If $\mathbf{e}$ and $\mathbf{b}$ denote time-dependent \textit{real} electric and magnetic 3-vector fields associated with any pulse solution $F_{\mu\nu} $, its total electromagnetic energy  $\J$, for a fixed set of parameters  and any $z$,  is calculated from
\begin{eqnarray}\label{P}
	\J &=& \frac{1}{\mu_0} \int_{-\infty}^{\infty} \, dt \, \int_{S}  (  { \mathbf e} \times {\mathbf b}   ) \cdot d{\mathbf S}
\end{eqnarray}
where $S$ can be any plane with constant $z=z_0>0$. For {\it spatially compact} pulse fields in vacuo this coincides with the total pulse electromagnetic energy
\begin{eqnarray}\label{ENERGY}
	{\cal E} &=& \int_{\cal V}  \rho\, d{\cal V} \,=\, \int_{-\infty}^{\infty}\!\! dz\,\int_0^{ 2 \pi} \!\!d\theta\, \int_{ 0} ^\infty\!\! r dr\;\rho(t,r,z,\theta)
\end{eqnarray}
where $\rho\equiv \frac{1}{2}\,\left( \epsilon_0\, {\mathbf e} \cdot {\mathbf e}  + \frac{ {\mathbf b} \cdot {\mathbf b}}{\mu_0 } \right)$ is integrated over all space ${\cal V}$. This follows since $\nabla\cdot({ \mathbf e} \times { \mathbf b} ) = -\mu_0\,\partial_t\,\rho $. To correlate $\J$ with other laser pulse properties and the choice of parameters, we bring the pulses $F^{s,\kappa}_{\mu\nu}$ for various values of $s$ and $\kappa$ into classical interaction with one or more charged point particles. The world-line in spacetime of a single particle, parameterized in arbitrary coordinates as $x^\mu= \xi^\mu(\tau)$ with a parameter $\tau$, is taken as a solution of the coupled non-linear differential equations
\begin{eqnarray} \label{EOM}
	{ { \cal A}_\mu  (\tau) } &=& \frac{q}{m_0 c^2}\FF_{\mu\nu}( \frac{}{} \!\xi(\tau) ) \, V^\nu(\tau)
\end{eqnarray}
in terms of the particle charge $q$ and rest mass $m_0$, for some initial conditions $\xi(0), V(0)$, where the particle 4-velocity satisfies $V^\nu\, V_\nu=-1$ and its 4-acceleration is expressed in terms of the Christoffel symbols $\Gamma^{\delta}{}^\beta{}_{\mu}$ as  ${  \cal A}_\mu(\tau) = \partial_\tau \, {V_\mu}(\tau) +  V_{\delta} (\tau)\,V_{\beta}(\tau)\, \Gamma^{\delta}{}^\beta{}_{\mu}( \xi( \tau) )$.  In the following, radiation reaction and inter-particle forces are assumed negligible. From the solution $\xi(\tau)$ one can determine the increase (or decrease) in  the relativistic kinetic energy transferred from the electromagnetic pulse to any  particle and the nature of its trajectory in the laboratory frame for different choices of $s$ and $\kappa$.  

\section{Interactions of Compact Laser Pulses with Matter}
The analysis of the previous section can be used to investigate the behaviour of a classical laser pulse interacting with electrically charged matter and motivates a model for a \textit{quantum laser pulse} interacting with electrically neutral matter. 

In order to facilitate the classical behaviour, we reduce the above equations of motion to dimensionless form and fix the physical dimensions of the fields involved. The Minkowski metric tensor field $g=g_{ \mu\nu} dx^\mu\, dx^\nu$ (with $g_{ \mu\nu }=\mathrm{ diag } (-1,1,1,1)$)  in inertial coordinates $x^0=c t,x^1=x,x^2=y,x^3=z$) has SI physical dimensions $[L]^2$. The SI dimension of electromagnetic quantities follows by assigning to $ \epsilon_0 F_{\mu\nu}\, dx^{[\mu} dx^{\nu]}$ in any coordinate system the physical dimension of charge. Furthermore, in terms of Minkowski polar coordinates $\{ t,r,z,\theta \}$, introduce (for ease of graphical visualization) the dimensionless coordinates $\{R=r/\Phi\W, \, T=c t/\W, \, Z=z/\Xi\W\}$ and real dimensionless parameters $\Lambda, \Psi_j=\psi_j/\W$ ($j=1,2$) where ${[\Psi_j]= [\Phi]=[\Xi]=1},{[\W]=[L]}$. Then with the dimensionless complex scalar field  $\widehat{\alpha }(T,R,Z)=\alpha(t,r,z)$  and Greek indices ranging over $\{ T, R, Z,\theta \}  $ with $ \epsilon^{ T,R,Z,\theta  } = 1$ ,  we write
\begin{eqnarray} \label{dimless}
	A_\delta &=& \frac{m_0 c^2 \W^3}{q} \,\,R\Lambda\,\partial_\gamma\left(\widehat{ \alpha} \, \widehat{ \Pi}_{\mu\nu }\right) \epsilon^{\gamma\mu\nu}{}_\delta 
\end{eqnarray}
for a choice of dimensionless covariantly constant tensor $\widehat{\Pi}_{\mu\nu }(T,R,Z,\theta)=\Pi_{\mu\nu}(t,r,z,\theta)$, so that $ [{\epsilon_0}A_\mu\, dx^\mu]$  has the physical dimension of electric charge. The total power density $P$ and total energy density $\hat{\cal E}$ are now defined by
\begin{eqnarray*}
	\J 		 &=& \int_{-\infty}^\infty\,dT\,\int_0^\infty\,d R\,\int_0^{ 2\pi}\,d\theta\,P(T,R,Z,\theta) \\[0.2cm]
	{\cal E} &=& \int_{-\infty}^\infty \,dZ\,\hat{ \cal E}(T,Z).
\end{eqnarray*}
The real parameter $\Lambda$ controls the strength of the electric and magnetic fields in $F_{\mu\nu}^{s,\kappa}$  for fixed values of the real parameters $\Psi_1,\Psi_2,\Phi,\Xi$ and the overall scale $\W$ will be fixed in terms of the total electromagnetic energy of the pulse. For a choice of such  parameters the associated  real fields $ {\mathbf e }$ and ${\mathbf b } $ enable one to calculate a numerical value $\GAM$ such that $\J=\W\GAM$. The diffraction of the pulse peak along the $z-$axis can be used to define a pulse range relative to the maximum of the pulse peak at $z=0$. To this end, the density $\EE( T,Z )$ defines the dimensionless range $Z_{rg}$ by ${\EE(0,0  ) }/{\EE( T_1,Z_{rg} )} = {2}$, where the peak at $Z=Z_{rg} > 0$ and $T=T_1 >0 $ is half the height of the peak at $Z=0, T=0$. If during the interval $[0,T_1]$ the pulse propagates with negligible deformation in $Z$, one may estimate its width $Z_w$ at half height and the dimensionless pulse axial speed $\beta=Z_{rg}/T_1$. This yields the dimensionless pulse duration or temporal width $T_0={Z_w}/{\beta}$. From these dimensionless values one deduces the pulse SI characteristics in terms of $\W$ and hence $\J$. If the picosecond is used as a unit of time, the pulse duration becomes $t_0= \W T_0/c= \W Z_w/(\beta c)=N\, 10^{-12}$ s for some value $N$ and hence $\W=(c \beta N/Z_w)\, 10^{-12}$ m, $\J= (\GAM {\beta }c N/Z_w) \, 10^{-12}$  J,  $z_w=\Xi c \beta n\,10^{-12} $ m and $z_{rg}= \W\Xi Z_{rg}=(\Xi{\beta} c N Z_{rg}/Z_w)\,10^{-12}  $ m. A dimensionless spot-size of the pulse at $ Z=Z_0 >0 ,\, T=Z_0/\beta $ is  then determined by  the behaviour of $ P(R,Z_0/\beta, Z_0,\theta)$.  At each value of $Z_0$ this function of $R$ and $\theta$ has a clearly defined principal maximum. If one associates a circle of dimensionless radius $R_s(Z_0)$ with such a maximum locus it can be used to define a spot-size at $z=z_0$ with radius $r_s(z_0)\!=\!\W\Phi R_s(Z_0)\!=\! (c \beta N \Phi R_s(Z_0)/Z_w)10^{-12}$ m.
\LASERTABLE
Figure~\ref{LPPD} displays a clearly pronounced principle maximum in the power density  profile $P$ as a function of $X=R\cos(\theta)$ and $Y=R\sin(\theta)$ at $Z=0,\,T=0$ for a specific choice of the parameters \params. The same parameter set is used to numerically solve (\ref{EOM}) for a collection of  trajectories for charged particles, each arranged initially around the circumference of a circle in a plane orthogonal to the propagation axis of incident CM type laser pulses with different chirality $\kappa$. The resulting space curves in 3-dimensions, displayed in figure~\ref{DTP}, clearly exhibit the different characteristic responses to CM pulses with distinct chirality values.
\LASERSKE
 The instantaneous {\it specific} relativistic kinetic energy of a particle with laboratory speed $v$ is $\gamma - 1$ in terms of the Lorentz factor $\gamma$ given by $\gamma^{-1}=\sqrt{1-\frac{v^2}{c^2}}$. 
In figure~\ref{SKE}, this quantity is displayed as a function of $T$ on the left for a charged particle accelerated by a fixed chirality (CM,$-1$) type pulse where the pulse energy is varied by changing $\Lambda$. On the right the energy transfer dependence on pulse chirality for both CE and CM type pulses with  {\it fixed laser energy} is displayed. We deduce that the pulse momentum and angular momentum \cite{allen2003optical} in the propagation direction can transfer an impulsive force and torque respectively to charges lying in an orthogonal plane. More generally, the classical configurations of a high energy pulse labelled CE or CM could be distinguished experimentally by their interaction with different arrangements of charged matter.

Furthermore, by a suitable choice of parameters, (CE, $\kappa$) type modes can be constructed that yield the same physical properties ($\J,z_{rg}, z_w,\beta$) for all $\kappa$. Similarly the (CM,$\kappa$) type modes yield a $\kappa$ independent set with physical properties distinct from those determined by the (CE,$\kappa$) modes. The pulse  group speed magnitudes (as defined above) of all these configurations are determined numerically and are bounded above by the value $c$. To illustrate some of these statements, Table~\ref{MKSLPC} summarizes the SI laser pulse characteristics determined by solving the system of equations in (\ref{EOM}) for a specific choice of $\{\Psi_1,\Psi_2, \Phi,\Xi\}\,$ and various values of $\Lambda$, selected to demonstrate the physical characteristics of existing laser pulses. We conclude that both the CE and CM type pulses can propagate localized packets with linear and angular  momentum analogous to the quantum characteristics of propagating wave packets describing particles with different masses.

These observations suggest that, for electromagnetic micropulses with ${\cal{E}} t_{0}\lesssim \hbar$, a parameterized {\it effective quantum Hamiltonian} $H$ describing a particular pair of interacting massive point particles may provide the simplest \textit{effective} description of general non-stationary quantum states of a laser pulse. Such states are then defined as elements of a complex Hilbert space $\HH=L^{2}(\real^{3},\complex^{3}) \otimes  L^{2}(\real^{3},\complex^{3})$ where each factor denotes the space of complex square integrable 3-component vectors on $\real^{3}$ carrying irreducible finite dimensional representations of the 3-dimensional rotation group\footnote{The inner product on $\HH=\HH_{1} \otimes \HH_{2}$ is defined by \unexpanded{$\langle \phi_{1} \otimes \phi_{2}, \phi'_{1} \otimes \phi'_{2} \rangle = \langle \phi_{1},\phi'_{1} \rangle_{\HH_{1}} \, \langle \phi_{2},\phi'_{2} \rangle_{\HH_{2}}$} for $\phi_{n}, \phi'_{n} \in \HH_{n}, \,n=1,2    
$  in terms of the inner products    $ \langle \phi_n,\phi'_n\rangle_{\HH_{n} }$  on $\HH_{n} $}. Such a construction is analogous to the Hilbert space $ \HH_{\text{vib}} \otimes \HH_{\text{rot}} $  used in Bohm and Loewe's dynamic model of a vibrating and rotating diatomic molecule after removing its translational degrees of freedom \cite{bohm1986quantum}. The space $\HH_{\text{vib}}$ is spanned by eigenstates of the Hamiltonian $ H_{\text{vib}} $ for a one-dimensional simple harmonic oscillator parameterized by a fundamental frequency $\omega_0$  and the space
$\HH_{\text{rot}}$ is spanned by eigenstates of the Hamiltonian $ H_{\text{rot}} $ for a one-dimensional rigid rotator parameterized by a constant moment of inertia parameter $\cal{I}$. The simplest  model for the diatomic molecule takes for its Hamiltonian:
\begin{eqnarray*}
	H_{\text{mol}} &=& H_{\text{vib}} \otimes I + I \otimes H_{\text{rot }} + {\cal G} W_{\text{vib}} \otimes W_{\text{rot}}  
\end{eqnarray*}
in terms of the three phenomenological real parameters $\omega_0, {\cal{I,G}} $ and where $I$ denotes the identity operator on the relevant space. For certain diatomic molecules this idealized model (and its refinements)  accounts for well-established empirical formulae for their electronic energy states. States in the space $\HH$ must describe the translational modes of a quantum packet as well as possible mode-mixing. To this end we model their unitary evolution generated by an effective Hermitian Hamiltonian of the form:
\begin{eqnarray*}
	H &=& H_{\text{0}}( {\mathbf x_{1}}  , {\mathbf x_{2}}   )   + H_{\text{int}}( {\mathbf x_{1}}  , {\mathbf x_{2}}   )   
\end{eqnarray*}
where
\begin{eqnarray*}
	H_{\text{0}} &=& \left(-\frac{\hbar^{2}}{2\mu_{1}}\nabla^{2}_{\mathbf x_{1}}\right) \otimes I + I \otimes \left(-\frac{\hbar^{2}}{2\mu_{2}}\nabla^{2}_{\mathbf x_{2}}\right)
\end{eqnarray*}
with real parameters $\mu_{1}, \mu_{2} >0$. For \textit{quantum} pulses that are deemed relativistic, one can include relativistic corrections  by the replacement
\begin{eqnarray*}
	-\frac{\hbar^{2}}{2\mu_{n}}\nabla^{2}_{\mathbf x_{n}} \quad\longrightarrow\quad \sqrt{-\hbar^{2}c^2 \nabla^{2}_{\mathbf x_{n}} + \mu_n^2 c^4 \;} -\mu_n c^2
\end{eqnarray*}
for $n=1,2$ and working in a momentum representation. The structure of  $H_{\text{int}}( {\mathbf x_{1}}  , {\mathbf x_{2}}   ) $   depends upon the physical properties of the medium through which the pulse propagates. A simple model that accommodates quantum {\it birefringence}, medium anisotropy and inhomogeneity may be constructed in terms of parameterized Hermitian operators $\WW_{1}(\mathbf x_{1}), \, \WW_{2}(\mathbf x_{2})  $:
\begin{eqnarray*}
	H_{\text{int}} &=& \WW_{1}(\mathbf x_{1}) \otimes \WW_{2}(\mathbf x_{2}).
\end{eqnarray*}
The six classical chiral states labelled $(s,\kappa)$ that evolve according to the classical  Maxwell equations are now replaced by  elements $\Psi_{t} \in \HH$ satisfying the Schr\"{o}dinger equation:
\begin{eqnarray*}
	-\frac{\hbar}{i}\frac{\partial}{\partial t}\Psi_{t} &=& H \Psi_{t}
\end{eqnarray*}
with $\Psi_{t_{0}}$ prescribed at any time $t_{0}$ and satisfying $\langle \Psi_{t_{0}} , \Psi_{t_{0}}  \rangle = 1$. In a direct product matrix representation, decomposable states in $\HH$ can be written $\bfalp\bfx{1}\otimes \bfbta\bfx{2}$ with
\begin{eqnarray*}
	\bfalp \bfx{1} &=& \THREEVEC{\alp{1}\bfx{1}}{\alp{2}\bfx{1}}{\alp{3}\bfx{1}} \qquad\text{and}\qquad 
	\bfbta \bfx{2} \;=\; \THREEVEC{\bta{1}\bfx{2}}{\bta{2}\bfx{2}}{\bta{3}\bfx{2}}.	
\end{eqnarray*}
By analogy with the description of non-relativistic spin$-\frac{1}{2} $ qubit states, call $\bfalp$ the left qutrit component of such a bi-qutrit state and $\bfbta$ the corresponding right qutrit component. The dynamic description of a general  state $\psi_t \in \HH$
requires an explicit interaction Hamiltonian operator. To model interactions with a classical  medium we envisage here a smooth fabricated meta-material with specified inhomogeneous and anisotropic characteristics. This is similar to the specification of the magnetic moment interaction of the qubit states (due to an unpaired  bound s-wave electron) of a  silver atom with a classical (inhomogeneous) static magnetic field in the Stern-Gerlach experimental arrangement. An important practical distinction however arises since the laser states, unlike electrons,  are electrically and magnetically neutral  and hence their interaction with electromagnetically neutral properties of a fabricated meta-material may be more versatile than in the Stern-Gerlach situation \cite{mott1929scattering}.

Consider, for example, the evolution of the left qutrit component $\bfalp\bfx1$ under the Hamiltonian:
\begin{eqnarray*}
	-\frac{\hbar^2}{2 \mu_1}  \nabla^2_{\mathbf x_{1}} \otimes I + G_1 {\boldsymbol S} \cdot {\boldsymbol N}\bfx{1}\otimes I
\end{eqnarray*}
where the constant $(3\times 3)$ Hermitian matrix-valued vector  ${\boldsymbol S}$ satisfies the commutation relation ${\boldsymbol S} \times {\boldsymbol S}=i {\boldsymbol S} , \, $
the 3-vector ${\boldsymbol N}\bfx{1}$ is a classical field of unit-vectors on some fabricated meta-material and $G_{1}$ is  a parameter with the physical dimensions of energy.  Such a Hamiltonian assigns a single preferred field of spatial directions ${\boldsymbol N}\bfx{1}$ for the interaction of quantum states with the medium. More complex media  could involve multiple anisotropies described by a multi-directional {\it set of unit-vector}  fields. 

Given ${\boldsymbol  \alpha}(0,{ \mathbf x_1 })$ with 
\begin{eqnarray*}
	\sum_{j=1}^3\int_V \vert \alpha_j(0,{\mathbf x_1})\vert^2 \,d^3{\mathbf x_1} &=& 1
\end{eqnarray*}
(assuming an unbounded medium $V$), one solves the system:
\begin{eqnarray*}
	-\frac{\hbar}{i}\frac{\partial \bfalp\bfx1}{\partial t} &=& -\frac{ \hbar^2}{2\mu_1 } \nabla^2_{\mathbf x_1}\bfalp\bfx{1} + G_1\, {\boldsymbol S} \cdot {\boldsymbol N}\bfx{1}\, \bfalp\bfx{1} 
\end{eqnarray*}
for the left qutrit component at $t>0$.
Multi-component wave-packet solutions can be constructed by  the standard Fourier transform methods used to construct  wave-packets for scalar fields. Thus in cylindrical polar coordinates $(r,\phi, z)$ defined with respect to a preferred {\it time-independent, fixed $z-$direction}  ${\boldsymbol N}=(0,0,1)$ in a {\it uniformly homogeneous} medium, wave-packet solutions to the above equation take the form:

\begin{eqnarray*}
	\alpha_{j}(t,r,\phi,z) &=& \!\!\sum_{m=-\infty}^\infty\int_{-\infty}^\infty \!\!dk\,A_{j,m}(k,s)\,\int_0^\infty \!\!\sqrt{s r}\,ds\,J_m(r s)\, \exp[ikz +im\phi]\;\Gamma_{j}(t,k,s)
\end{eqnarray*}
for $j=1,2,3$ where 
\begin{eqnarray*}
	\Gamma_{1}(t,k,s) &=& \exp\left[ \frac{it}{\hbar}\left(\frac{\hbar^2}{2 \mu_1} ( k^2 + s^2  ) + G_{1}\right) \right] \\
	\Gamma_{2}(t,k,s) &=& \exp\left[ \frac{it}{\hbar}\left(\frac{\hbar^2}{2 \mu_1} ( k^2 + s^2  ) \right) \right] \\
	\Gamma_{3}(t,k,s) &=& \exp\left[ \frac{it}{\hbar}\left(\frac{\hbar^2}{2 \mu_1} ( k^2 + s^2  ) - G_{1}\right) \right] .
\end{eqnarray*}
and the amplitudes $ A_{j,m}(k,s)$ are determined from the normalised initial conditions $\alpha_j(0,r,\phi,z)$ by Fourier and Fourier-Bessel inversion. In these expressions contributions to $\alpha_j$ from terms in the sums that depend on the integer $m$ indicate those from eigenstates of the orbital angular momentum operator $  \frac{\hbar}{i}{ \mathcal L } _{ \partial_\phi}$ \cite{matula2013atomic}. By contrast, terms in $\bfalp$ with different $j$ refer to spin (qutrit) contributions to $\psi_{t}$.  With the above Hamiltonian, stationary qutrit energy eigenstates would be non-degenerate with energy shifts $ (0,\pm G_1) $.    Generating superpositions involving left and right qutrit states in $\HH$ that cannot be reduced to decomposable states by a change of basis may offer a  means to  isolate and thereby control non-stationary entangled qutrits using appropriately  fabricated meta-materials.

In addition to such  laser  state \lq\lq measurement\rq\rq \,interactions with a classical medium, one may also include interactions with atomic quantum states. These and other uses of such effective quantum Hamiltonians involving the dynamics of laser states will be discussed elsewhere.


\newpage
\section{Conclusions}
We have constructed a basis of classical  chiral solutions of the source-free vacuum Maxwell field equations  from a  simple particular solution to the complex scalar wave equation in spacetime and a set of covariantly constant antisymmetric tensor fields.  Such solutions offer a simple 3-parameter description of a finite-energy laser pulse that provide a more accurate simulation tool for analysing laser-matter interactions in realistic 3-dimensional situations where plane-fronted paraxial approximations are inadequate.  The analytic structure of such solutions enables one to readily extract all the standard diffractive characteristics associated with a laser pulse in free space.
Using the classical relativistic Lorentz-force equation of motion we have also analysed numerically the interaction of such compact pulse solutions with charged point particles. This has  explicitly demonstrated how laser configurations with definite chirality and mode-type transfer angular momentum and energy to the charges as a result of the interaction. From these numerical investigations we have proposed a particular effective quantum model for systems where the classical pulse energy ${\cal{E}}$  and the pulse duration $t_{0}$  satisfy ${\cal{E}} t_{0}\lesssim \hbar$. By analogy with the effective modelling of rotating-vibrating diatomic molecules we have proposed a simple phenomenological Hamiltonian that may be used to describe  quantum laser packets in free space and material media. It is suggested that this Hamiltonian may have utility for simulating  a novel transfer of quantum information and for constructing  models of rapid single-cycle laser pulses  interacting with quantum matter and classical fabricated materials containing structures below the nano-scale.

\section*{Acknowledgements}
	The authors RWT and TW are grateful to STFC (ST/G008248/1) and EPSRC \\ (EP/J018171/1) for support.  SG gratefully acknowledges support from Grant-in-Aid for Young Scientists (Grant No. 25800181). The authors acknowledge useful discussions with Dino Jaroszynski, Robert Young and other colleagues in the ALPHA-X project, the Cockcroft Institute and Lancaster University.

\newpage
\bibliographystyle{unsrt}
\bibliography{Laser_JPHYSA}

\end{document}